\documentclass[referee]{aastex}
\usepackage{multirow}

\begin{document}

\title{Star formation properties in the Local Volume galaxies via $H\alpha$ and FUV fluxes.}
\author{Igor D. Karachentsev \altaffilmark{1} and Elena I. Kaisina}
\affil{Special Astrophysical Observatory RAS,  Nizhnij Arkhyz,    Karachai-Cherkessian Republic,  Russia 369167}
\altaffiltext{1}{Leibniz-Institute fur Astrophysik (AIP), An der Sternwarte 16, D-14482, Potsdam, Germany}
 \email{ikar@sao.ru,kei@sao.ru}

\begin{abstract}
 A distance-limited sample of 869 objects from the Updated Nearby Galaxy
Catalog is used to characterize the star formation status of the Local
Volume population. We present a compiled list of 1217  star
formation rate (SFR) estimates for 802 galaxies within 11 Mpc from us, derived from the
H-alpha imaging surveys and  GALEX far-ultraviolet survey. We briefly
discuss some basic scaling relations between SFR and luminosity, morphology,
HI-mass, surface brightness, as well as environment of the galaxies. About
3/4 of our sample consist of dwarf galaxies, for which we offer a more refined
classification. We note that the specific star formation rate of nearly
all luminous and dwarf galaxies does not exceed the maximum value:
$\log(SFR/L_K) = -9.4$ [yr$^{-1}$]. The bulk of spiral and blue dwarf galaxies
have enough time to generate their stellar mass during the cosmological time,
$T_0$, with the observed SFRs. They dispose of a sufficient amount of gas to
support their present SFRs over the next $T_0$ term. We note that only
a minor part of BCD, Im, and Ir galaxies ( about 1/20) proceeds in
a mode of vigorous star-burst activity. In general, the star formation
history of spiral and blue dwarf galaxies is mainly driven  by their internal
processes. The present SFRs of E, S0 and dSph galaxies are typically
(1/30 - 1/300) of their former activity.

\end{abstract}

\section{Introduction}

A  series of surveys of nearby galaxies in the  $H\alpha$ emission line,  determining the  star formation characteristics in them was published
over the last decade. Typically, the object of study were the galaxies of
a fixed morphological type: blue compact galaxies, BCD (Gil de Paz et al. 2003), irregular and
BCD (Hunter \& Elmegreen, 2004), spiral and irregular galaxies (James et al. 2004, Epinat et al. 2008),  southern objects, rich in neutral
hydrogen HI (Meurer et al. 2006), dwarf galaxies in the nearby Sculptor  and Centaurus~A groups ( Bouchard et al. 2009).

The most systematic observations in the  $H\alpha$ line were made by Kennicutt et al. (2008), who have selected for their survey
the nearby ($D < 11$ Mpc) galaxies with the apparent magnitude $B<15^m$ at the Galactic latitudes $\mid b\mid > 20^{\circ}$
 with the morphological types T $> -1$ in de Vaucouleurs classification. In parallel with this, a program of mass survey of  galaxies of
the Local Volume ( = LV) without any object selection by
morphological type  was performed at the 6-m BTA telescope of the
Russian Academy of Sciences (Karachentsev et al. 2005, Kaisin \&
Karachentsev 2006, 2008, Kaisin et al. 2007, 2011, Karachentsev
\& Kaisin 2007, 2010). In total we have obtained the
$H\alpha$-images of more than 300 galaxies with distances of $D
<11$ Mpc within our  program. The summary of $H\alpha$ fluxes for
more than 500 LV-galaxies was published in the Updated Nearby
Galaxy Catalog (= UNGC) (Karachentsev et al. 2013a). The
database and atlas of the LV-galaxies (Kaisina et al. 2012),
available at  http://www.sao.ru/lv/lvgdb, contains a lot of
$H\alpha$-images for nearby galaxies, their $H\alpha$ fluxes
along with the flux uncertainties, as well as references to the
sources of the observables. Actually, our sample contains nearly
2 times more galaxies than the most recent study by Lee et al. (2011),
being also much more representative for the early-type objects (E, S0, dSph)
and for the low-mass galaxies with $\log(SFR) < -4$.

It should be emphasized that the sample constraint  by a fixed
distance and minimal sample selectivity during its compilation are
very important circumstances that facilitate interpretation of
the obtained data. For example, the use of our sample allows to
obtain a less biased  estimate of the average  star
formation rate (SFR) in a unit volume at the present epoch (z =
0). It is also useful as a reference  sample in the analysis of
the effects produced by  the overdensities in the nearby
Virgo and Fornax clusters on the features of star formation in
their galaxies.

An independent possibility of determining the star formation
rates in the nearby galaxies by their far-ultraviolet flux
$(\lambda_{eff}=1539$\AA, FWHM=269\AA)  has appeared with the FUV
flux measurements by the GALEX orbital telescope  (Gil de Paz et
al. 2007, Lee et al. 2009, 2011). 
The consolidated data of these authors was
significantly supplemented by us with integral
 $m_{FUV}$ magnitudes of galaxies extracted
from the NASA Extragalactic Database and then presented in the UNGC (Karachentsev et al. 2013a) and the LVG data base (Kaisina et al. 2012).
The sample discussed below is the most representative of all samples to date.

\section{The data sample}

The UNGC catalog includes 869 galaxies of the northern and
southern sky with individual distance estimates of $D<11$ Mpc or
radial velocities relative to the Local Group centroid of $V_{LG}
<600$ km s$^{-1}$ if the galaxy distance has not been determined.
More than 320 galaxies of the sample have distance estimates
derived with an accuracy of $\sim10$\% by the TRGB  luminosity,
the luminosity of Cepheids or Supernovae, or via the surface
brightness fluctuations in the galaxy. The distances of 370
members of the Local Volume were  determined with an accuracy of
$\sim25$\% from the Tully-Fisher (1977) relation, luminosity of
the brightest stars or evident   membership of galaxies in the
nearby groups. The kinematic distance estimates of nearby galaxies
$D=V_{LG}/H_0$, where  $H_0$   is the Hubble parameter are less
reliable, because they may contain a significant bias owing to
the  participation of nearby galaxies in the large-scale flows
from the  center of the Local Void or towards the Virgo cluster
(Tully et al. 2008).

Following Kennicutt (1998), we determined the integral SFR  in a galaxy by the linear relation
$$ SFR[M_{\odot}/yr]=0.945\cdot 10^9 F_c(H\alpha)\cdot D^2, \eqno(1)$$
where $F_c(H\alpha$) is its integral flux in the  $H\alpha$ line [erg$\cdot$cm$^{-2}\cdot$ s$^{-1}$],  corrected for
the Galactic and internal extinction  $A(H\alpha)=0.538(A^G_B+A^i_B)$.

The light extinction in our Galaxy  $A^G_B$ was estimated
according to Schlegel et al. (1998), and   internal extinction  $A^i_B$  was expressed through the apparent axial ratio of the galaxy $a/b$:
$$ A^i_B=[1.54+2.54(\log 2V_m-2.5)]\log(a/b), \eqno(2)$$
if the amplitude of the galaxy rotation $V_m$ exceeded 39 km s$^{-1}$ (Verheijen 2001);  for dwarf galaxies with  $V_m<39$ km s$^{-1}$ and for
the gas-poor E, S0 galaxies the internal extinction was assumed to be negligible.

According to Lee et al. (2011), the SFR in a galaxy can be expressed in terms of its apparent FUV magnitude $m_ {FUV} $ as
$$\log(SFR[M_{\odot}\cdot{\rm yr}^{-1}])=2.78-0.4m^c_{FUV}+2\log D \eqno(3)$$
accounting for the Galactic and internal extinction
$$m^c_{FUV}=m_{FUV}-1.93(A^G_B+A^i_B). \eqno(4)$$

Note that  Kennicutt \& Evans (2012) have recently provided a
little lower conversion factor between the SFR and the observed
$H\alpha$ or FUV flux, what is mainly the result of a different
initial mass function and updated stellar population models.

In total, our sample contains  619 galaxies with the  SFR values
determined from the FUV fluxes, and 98 galaxies with an upper
limit corresponding to $m_{FUV}\simeq23.0^m$. For 461 galaxies of
the Local Volume the SFRs are measured  from their  fluxes in
the  $H\alpha$ line, and for 41 galaxies only the upper limits of
their integral $H\alpha$ flux are known. Among both subsamples
there are 415 galaxies, the SFRs of which can be estimated by two
independent methods. This gives the so far largest  basis to
compare the methods used.

\section{Comparison of SFRs from $H\alpha$ and FUV fluxes}

\begin{figure*}
\centerline{
\begin{tabular}{c}\\[-25mm]
\includegraphics[width=0.6\textwidth,clip]{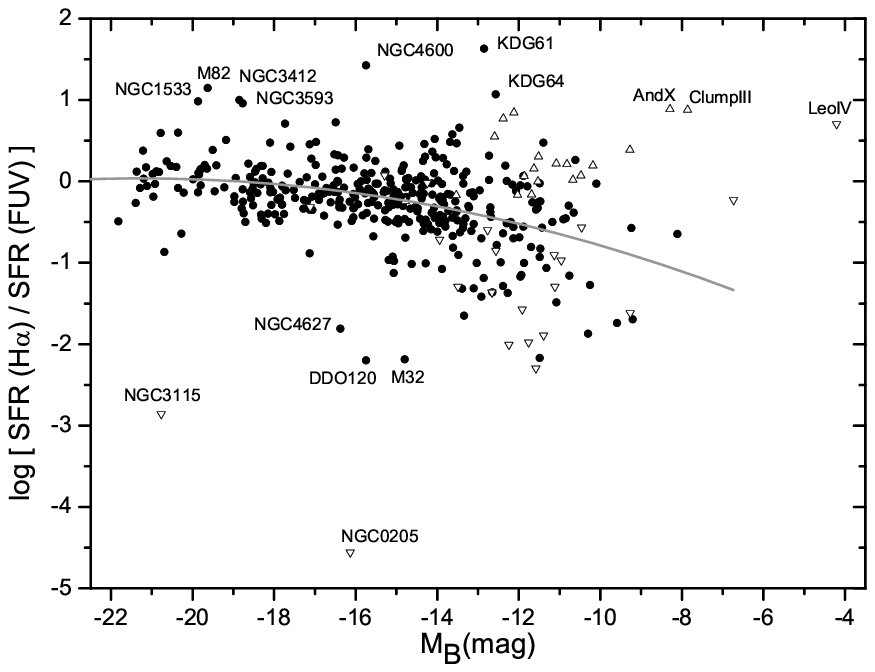}\\[-15mm]
\includegraphics[width=0.6\textwidth,clip]{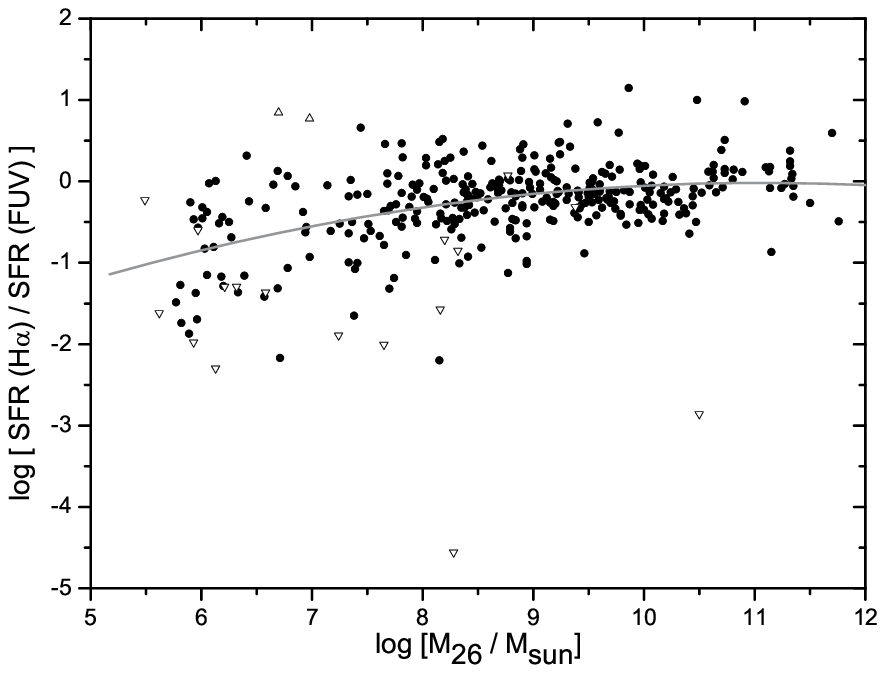}\\[-15mm]
\includegraphics[width=0.6\textwidth,clip]{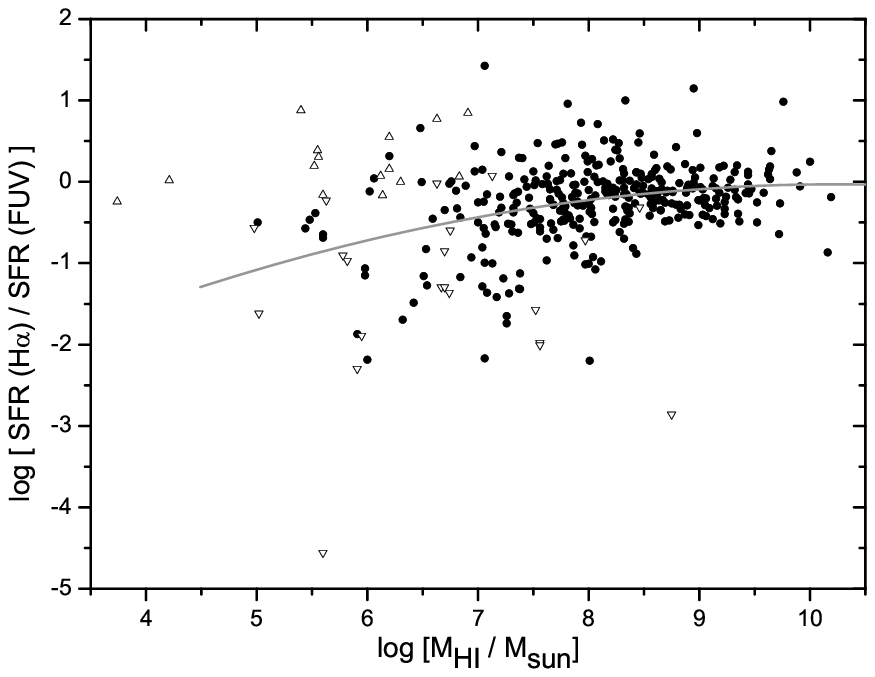}\\[-10mm]
\end{tabular}
}
\caption{The ratio of $H\alpha$-to-FUV star formation rates as a function
       of different global galaxy parameters: absolute blue magnitude
       (top panel), indicative dynamical mass within the Holmberg diameter
       (middle panel), and total hydrogen mass (bottom panel). The solid
       lines show the quadratic least squares fits to the data.}
\label{Fig1}
\end{figure*}

Figure~\ref{Fig1} represents the ratio of SFRs, determined via
$H\alpha$ and FUV fluxes depending on various global parameters
of the galaxies. The members of the Local Volume, for which  only
an upper limit of the flux in $H\alpha$ or FUV is known are shown
by empty triangles  pointing down or up, respectively. The
remaining objects are shown by circles. The top panel of the
figure shows the dependence of the SFR ratio on the absolute
B-magnitude of the galaxy. The middle panel gives the SFR ratio
as a function of the indicative dynamic mass $M_ {26}$ determined
within the Holmberg's isophote, 26.5 mag arcsec$^{-2}$. The
bottom panel shows the ratio of the SFRs as a function of the
total mass of hydrogen $M_{HI}$. The observed scatter of galaxies
in the diagrams of Fig. 1 is due to a variety of causes that are
both random and systematic.

First of all, a typical uncertainty of the measured
$H\alpha$ flux is around (10 - 20)\%, as it was noted by
Kennicutt et al. (2009) and Karachentsev \& Kaisin (2010). A
characteristic uncertainty of FUV fluxes derived from the GALEX
data is much lower, however, a scatter of FUV extinction in a
galaxy can reach about (20 - 40)\%, as shown by Lee et al.
(2009). If we also take into account   the distance measurement
errors (10 - 20)\%, one would expect a total uncertainty on
SFR($H\alpha$)-to-SFR(FUV) ratio to be somewhat within 50\%. But
Figure 1 demonstrates a much broader scatter.

Let us recall that the flux in the $H\alpha$ line determines the
SFR in a galaxy on a short time scale of $\sim10^7$  years
(because of the luminance of   O stars), while the FUV flux is
mainly formed by less massive stars of B0-B5 types and
corresponds to the time scale of $\sim10^8$ years. Due to the
bursts of star formation particularly significant in the most low-mass dwarfs (Stinson et
al. 2007, Skillman 2005), a scatter of the SFR ratios derived by
$H\alpha$ and FUV fluxes should increase with decreasing
luminosity or mass of the galaxy. This well expected effect is
actually observed in all the panels of Fig.1. The examples of
galaxies in central parts of which  the bursts of star formation
have occurred are M~82,  NGC~3412, NGC~3593, NGC~4600.

In the transition from  normal disk galaxies to diffuse dwarf
objects of low surface brightness, a relative error of
determining the $H\alpha$ flux usually increases. Here, an
underestimation of the integral $H\alpha$ emission  due to the
loss of the low-contrast component, distributed outside the
compact HII-regions can have a systematic effect. On the other
hand, the FUV-images of some diffuse dSph galaxies like the
Sculptor, AndI, AndII, AndXI, AndXVII may be contaminated by the
foreground blue stars of the Milky Way, what leads to a
fictitious increase of the FUV flux in the target galaxy. In
addition, the optical contours of dwarf galaxies  sometimes
contain poorly subtracted traces of very red stars, what
fictitiously increases the $H\alpha$ flux (probable examples:
AndIII, AndV, AndX). There are also rare cases when a large
difference in the SFR estimates from $H\alpha$ and FUV fluxes was
caused by small emission knots on the periphery of giant galaxies
that are accidentally projected onto the images of dwarf
satellites. Examples of such cases are  KDG61 as a companion to
M81 (Karachentsev et al. 2011) and M32 as a companion to M31.

Finally, we point out two instances:
 DDO~120 = UGC~7408 and  NGC~1533,  where significant errors in the measured $H\alpha$ flux are caused by a
poor control of weather conditions during the observations or
problems with data reduction.

In addition to all of the listed circumstances, it must be
remembered that the transition from the $H\alpha$ and FUV fluxes
to the SFR values is based on the  empirical relationships (1) --
(4), the validity of which is not entirely justified.
Pflamm-Altenburg et al. (2007, 2009) placed emphasis on this
cardinal problem, according to whom there is a systematic
underestimation of SFR in dwarf galaxies from their $H\alpha$
fluxes. The   authors see the cause of discrepancy  in the
features of the initial function of stellar mass in dwarf
galaxies (the deficiency of stars of the highest luminosity in
them). Detailed discussions on these issues can be found in 
Lee et al. (2009, 2011), Meurer et al. (2009), Hunter et al.
(2010), Fumagelli et al. (2011), Weisz et al. (2012), and Relano
et al. (2012).  In particular, Fumagelli et al. (2011) suggested
that a lower $H\alpha$-to-FUV flux ratio in dwarf galaxies
compared with that  in the brighter systems could be explained by
the stochastic sampling even for a universal initial mass
function. Weisz et al. (2012) considered the effects of a bursty
star formation history (SFH) and found a set of SFH models
that are well-matched with the observational data, implying that
the more massive galaxies have nearly constant SFHs, while the
low-mass systems experience burst amplitudes of 10 - 50. Their
bursty models are able to reproduce both the observed systematic
decline and an increased scatter in the $H\alpha$-to-FUV ratios
towards the low-mass systems.

\begin{figure*}
\centerline{
\begin{tabular}{c}\\[-20mm]
\includegraphics[width=0.9\textwidth,clip]{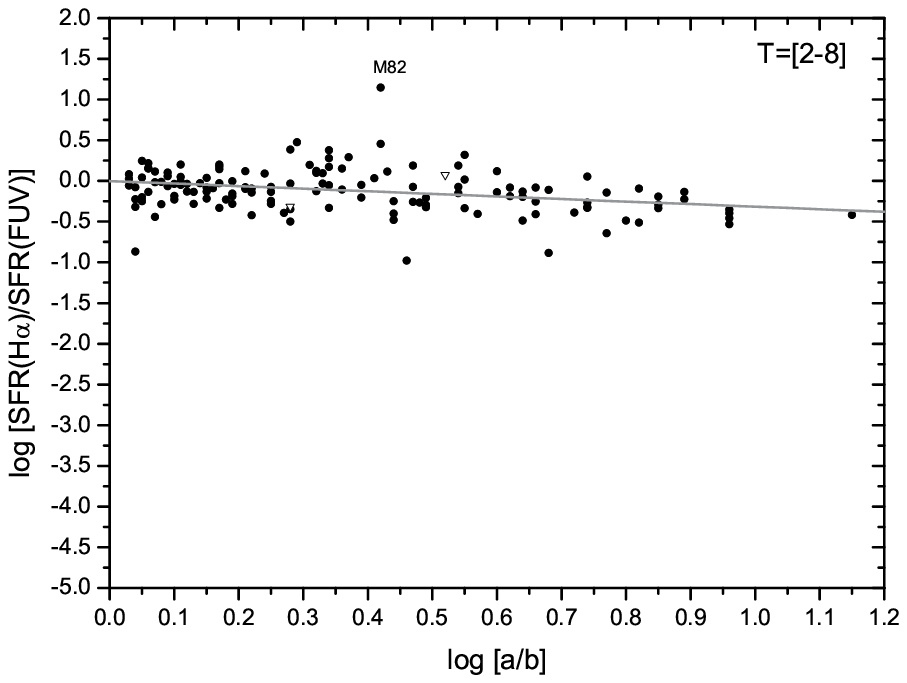}\\[-15mm]
\includegraphics[width=0.9\textwidth,clip]{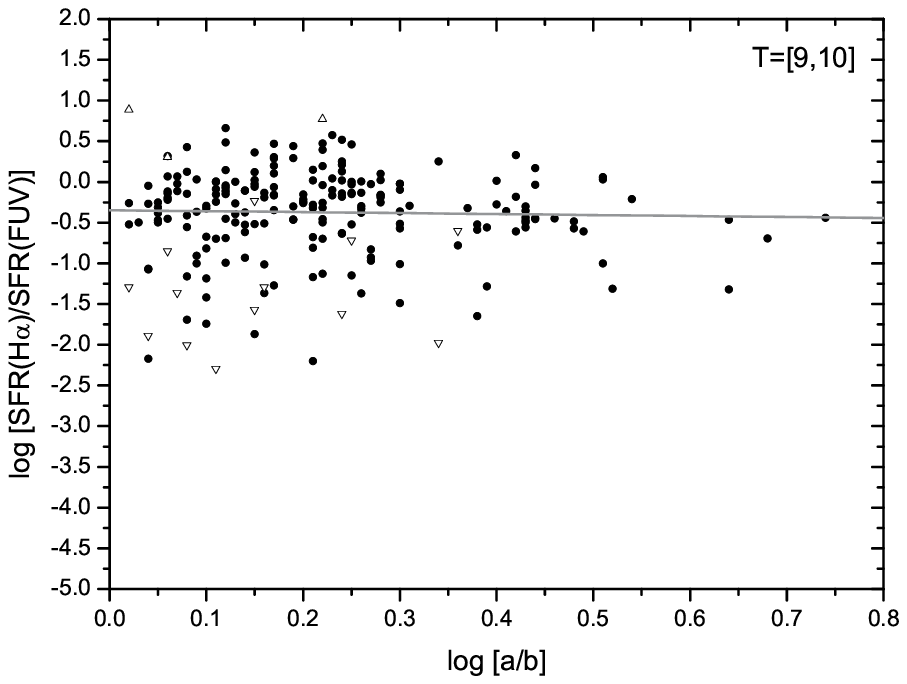}\\[-10mm]
\end{tabular}
}
\caption{The ratio of $H\alpha$-to-FUV SFRs vs. apparent axial ratio for
       spiral galaxies (top panel) and late-type irregular dwarfs (bottom
       panel). The solid lines indicate the linear least squares fits to
       the data.}
\label{Fig2}
\end{figure*}

One reason for the discrepancy between the SFR estimates from the
$H\alpha$ and FUV fluxes can be a wrong account of the internal
light extinction in the galaxies. To test this reason, we
compared the ratio of SFR estimates from $H\alpha$ and FUV fluxes
with an apparent axial ratio of galaxies $a/b$ in the logarithmic
scale. As the data in the upper panel of Fig.~\ref{Fig2} shows,
the disk spiral galaxies of   morphological types
 T=(2--8), i.e. T=(Sa-Sdm) have a relatively small scatter of
 $\log[SFR]_{H\alpha} - \log[SFR]_{FUV}$ with the average value being near zero.
 Consequently, the semi-empirical relations (1) and (3) are mutually well-calibrated for normal disk galaxies.
 Some trend of decreasing  $[SFR]_{H\alpha}/[SFR]_{FUV}$  towards the galaxies with large inclinations
 indicates a possible overestimation of the value of internal extinction in the disks by the relations (2) and (4).

 For the dwarf galaxies of T = 9, 10 morphological types  (BCD, Im, Ir), the logarithmic difference of SFR estimates
 is characterized by a significantly greater dispersion. On the average, the value of   $[SFR]_{FUV}$  is two times higher
 than   $[SFR]_{H\alpha}$  and is practically independent on the angle of inclination of galaxy.
 It is interesting to note that the scatter of difference of  $\log[SFR]$ shows a decreasing trend from the
 dwarf galaxies seen  face-on  to the edge-on ones.

\begin{figure*}
\centerline{
\begin{tabular}{c}\\[-20mm]
\includegraphics[width=0.9\textwidth,clip]{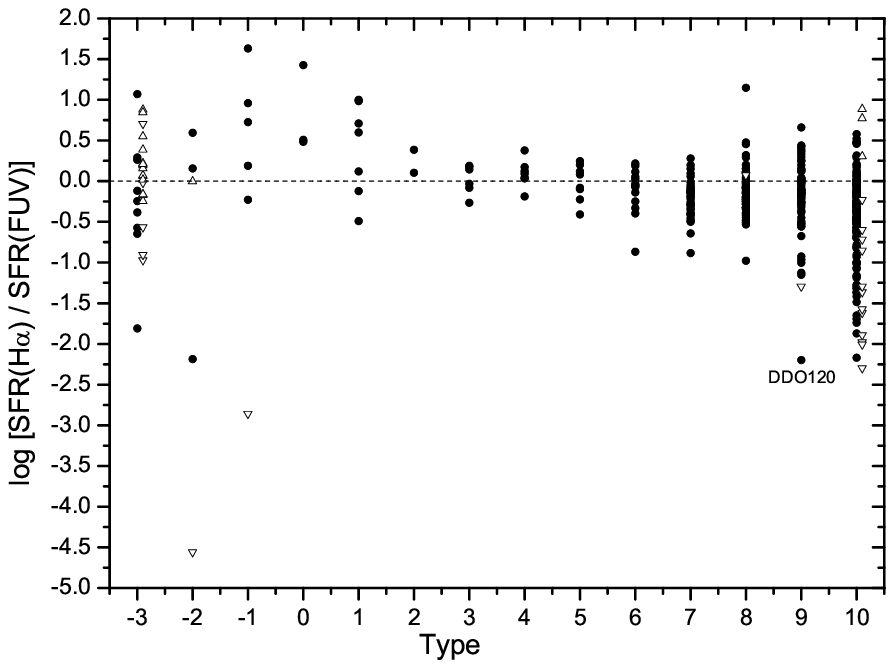}\\[-15mm]
\includegraphics[width=0.9\textwidth,clip]{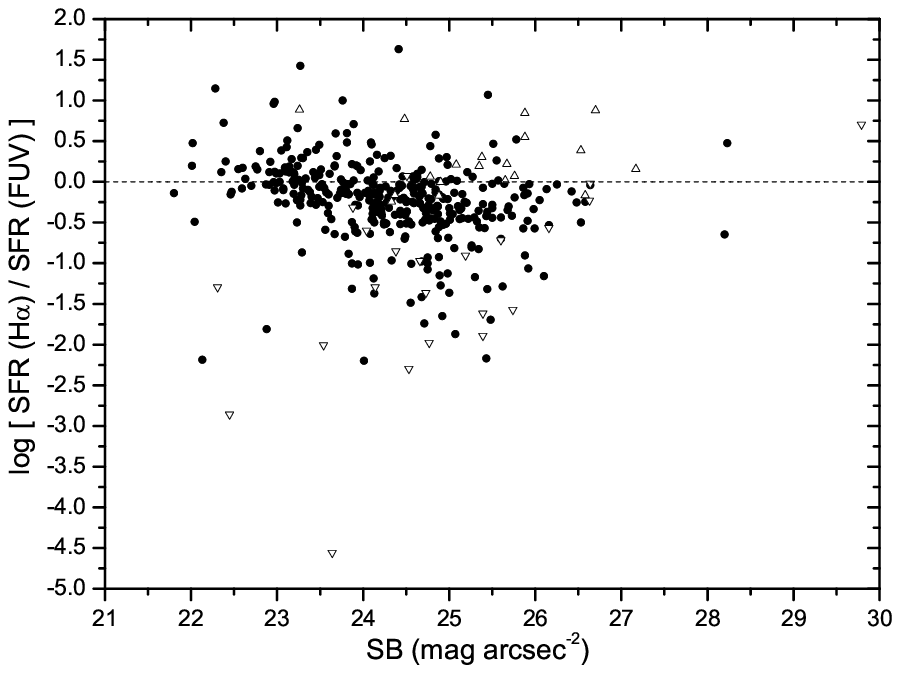}\\[-10mm]
\end{tabular}
}
\caption{The ratio of $H\alpha$-to-FUV SFRs vs. morphological type of galaxies
       on de Vaucouleurs scale upper panel) and vs. the mean B-band surface brightness within the Holmberg radius (bottom panel).
	   The galaxies with upper limit of $H\alpha$
       or FUV flux are shown by open triangles directed towards top or
       bottom, respectively.}
\label{Fig3}
\end{figure*}

 The difference    $\log[SFR]_{H\alpha} - \log[SFR]_{FUV}$,  depending on the morphological type of galaxies is 
shown in the upper panel of  Fig.~\ref{Fig3}.
As follows from these data,  for early-type (T$ <$2) galaxies,
the SFR estimates from  $H\alpha$ flux are on the average   2--5
times higher than those found from the FUV flux, also showing  a
large scatter here. It is not entirely clear for us what causes
this feature. An average difference between the estimates of
$\log[SFR]$ decreases from the early to late types, and their
variance is minimal for the spiral disks of T = (2--5),
systematically increasing towards the latest types. The reported
trends of the SFR estimates via H$\alpha$ and FUV fluxes along
the Hubble sequence have not yet obtained a physical
interpretation.

 Meurer et al. (2009) paid attention to the correlation between $H\alpha$-to-FUV flux ratio and a galaxy surface brightness.
 They found that low R- band surface brightness galaxies have lower ratio of the fluxes compared to high surface brightness galaxies. They
 presented this as the ``strongest'' evidence for systematic initial mass function variations, supporting the idea proposed by 
 Pflamm-Altenburg et al.(2009). On the bottom panel of Fig.~\ref{Fig3} we present the $H\alpha$-to-FUV SFR ratio versus B- band surface 
 brightness of a galaxy within its Holmberg radius. As one can see, our much more representative sample shows also the mentioned correlation,
 but only as a slight tendency. The observed noisy trend may be caused by known correlations of the mean surface brightness with other
 galaxy parameters: luminosity, morphology, HI-content, etc.

\section{Some scaling relations}

It is well known that  integral star formation rate in disk galaxies is approximately proportional to their integral luminosity,
i.e. the specific  star formation rate per  luminosity unit in them  is roughly one and the same (Young et al. 1996, Karachentsev \& Kaisin 2007,
James et al. 2008, Lee et al. 2009). This assertion is true, however, only in the first approximation. In addition to the luminosity, there \
apparently exist other parameters of galaxies affecting the difference in   specific star formation rates (SSFR).

\begin{figure*}
\centerline{
\begin{tabular}{c}\\[-25mm]
\includegraphics[width=0.6\textwidth,clip]{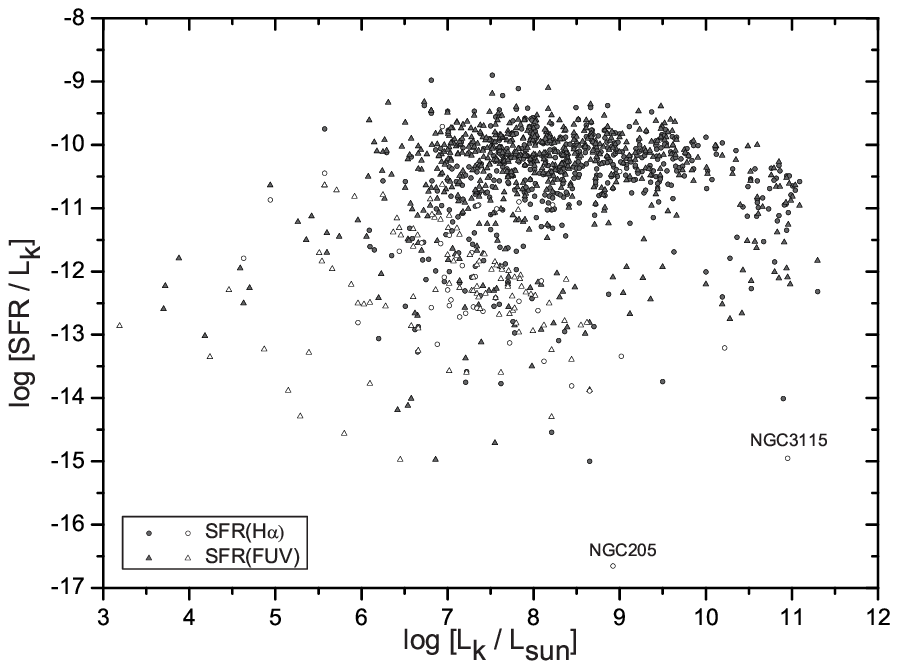}\\[-15mm]
\includegraphics[width=0.6\textwidth,clip]{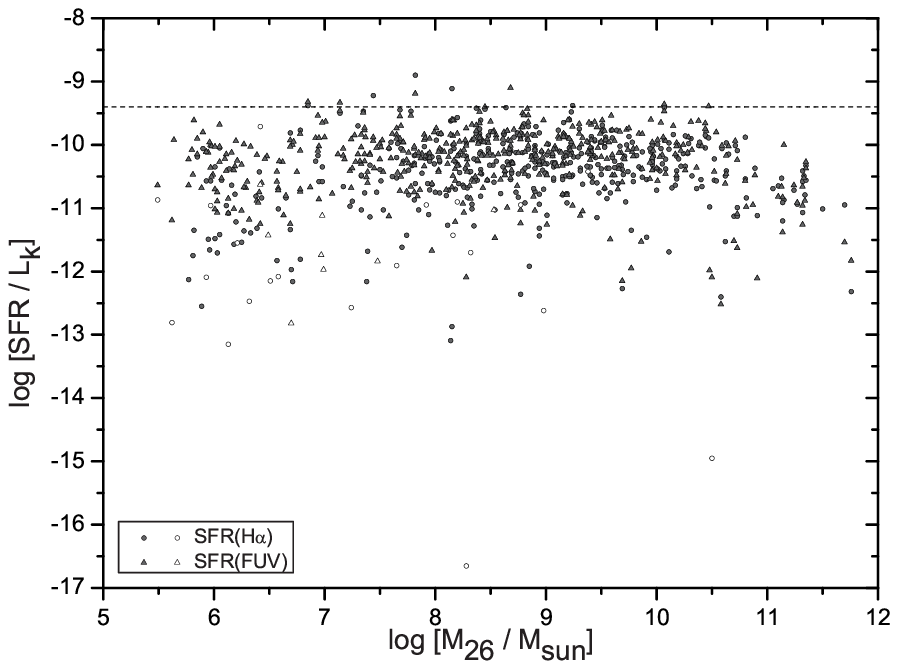}\\[-15mm]
\includegraphics[width=0.6\textwidth,clip]{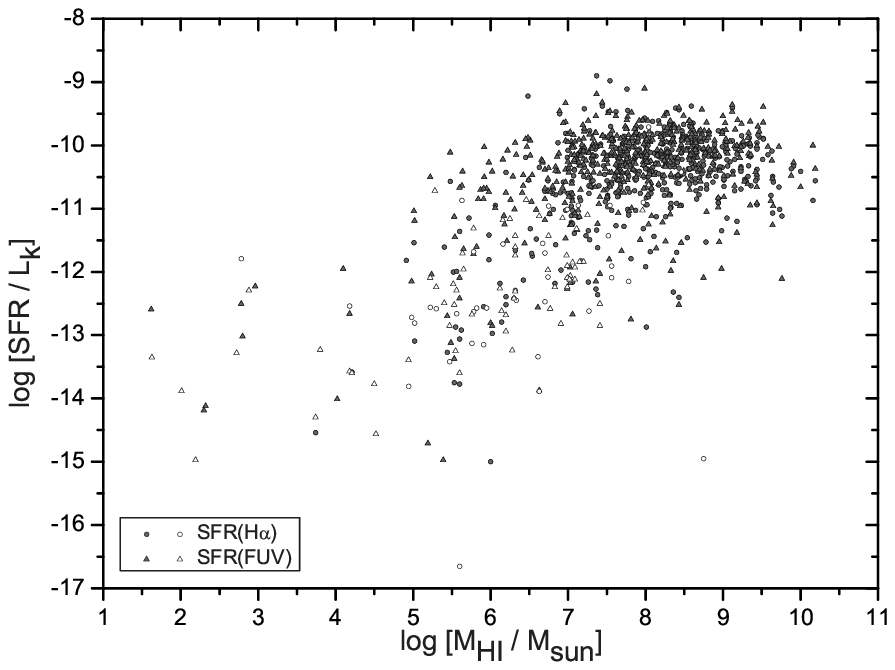}\\[-10mm]
\end{tabular}
} \caption{Specific star formation rates of nearby galaxies vs.
their different
       global parameters: $K_s$-band luminosity (top panel), dynamical mass
       within the Holmberg diameter (middle panel), and total hydrogen mass
       (bottom panel). Circles are based on $H\alpha$ fluxes, triangles are based
       on FUV fluxes; open symbols correspond to the upper flux limits.}
\label{Fig4}
\end{figure*}

The top panel of Fig.~\ref{Fig4} shows the distribution of
galaxies in the Local Volume by their SSFR per luminosity unit in
the $K_s$-band and the logarithm of  integral K-luminosity. The
galaxies with SFRs estimated from the $H\alpha$ and
 FUV fluxes are depicted by circles and triangles, respectively. Empty symbols denote galaxies with an upper limit of the $H\alpha$
 or FUV flux. Since  the stellar population of galaxies has an average mass-to-luminosity ratio in the K-band of  $\sim1 M_{\odot}/L_{\odot}$
 (Bell et al. 2003), the  $L_K$-scale actually matches the scale of integral stellar mass of the galaxies, $M_*$. In this diagram, the galaxies of
the Local Volume vary by 8 orders of magnitude in luminosity and
6 orders of magnitude in their SSFRs. In addition to the
horizontal \ "Main Sequence" for disk galaxies, a vertical
"column" stands out on the right side of the diagram, which
consists of  galaxies of the highest luminosity. Most of
them are early-type (Sa-Sb) spirals containing a prominent bulge
with old population. Different bulge-to-disc fractions in them
apparently lead to the observed scatter along the vertical scale.
Even greater differences in the specific SFR are revealed in the
dwarf galaxies. Only in a minor part  they are due to the increase
of $H\alpha$  and FUV flux measurement errors in the
low-luminosity galaxies near the sensitivity threshold of the
$H\alpha$ and FUV surveys. The main reasons of the observed
scatter are physical: the sweep-out of gas from the dwarf galaxies
during their passage through extended haloes of giant galaxies,
and cycles of star
 formation bursts which are the most significant in low-mass systems.

 It should be emphasized again that the distribution of galaxies in this diagram is subjected to selection effects only in a
 minimal degree, as compared with other similar samples.

The middle panel of Fig.~\ref{Fig4} shows the dependence of SSFR
on the dynamic mass of a galaxy, determined within the Holmberg
diameter by the amplitude of  internal motions $V_m$. The shape
of this diagram  greatly differs from the top one, since the
galaxies  poor in neutral hydrogen are underrepresented here. The
horizontal \ "Main Sequence" \  looks clearer here and shows the
presence of an upper boundary $\log(SFR/L_K)_{max}\simeq -9.4$,
above which there are just a few   extremely blue peculiar
galaxies:
  Garland, Mrk~209, Mrk~36, NGC~1592, UGCA~292.

Another parameter affecting the SFR   is the total hydrogen mass of galaxy $M_{HI}$. As follows from the bottom panel of Fig.~\ref{Fig4}, the most rapid
transformation of gas into stars occurs in the galaxies having large amount of neutral hydrogen. The slope of the log-log relationship between
the SSFR and  $M_{HI}$ in the region of    $\log(M_{HI}/M_{\odot})<7$ looks much steeper than in the galaxies with large hydrogen masses.
A different nature of  galaxy distribution in  three panels of Fig.~\ref{Fig4} again recalls
 that different conditions of galaxy selection in the considered sample  by optical or HI features can greatly influence the shape and
the subsequent interpretation of the observational data. This
fact was also noted by Huang et al. (2012) when they compared the
their
 samples, organized according to the ALFALFA, SDSS and GALEX survey data.

\begin{figure*}
\epsscale{1.0}
\plotone{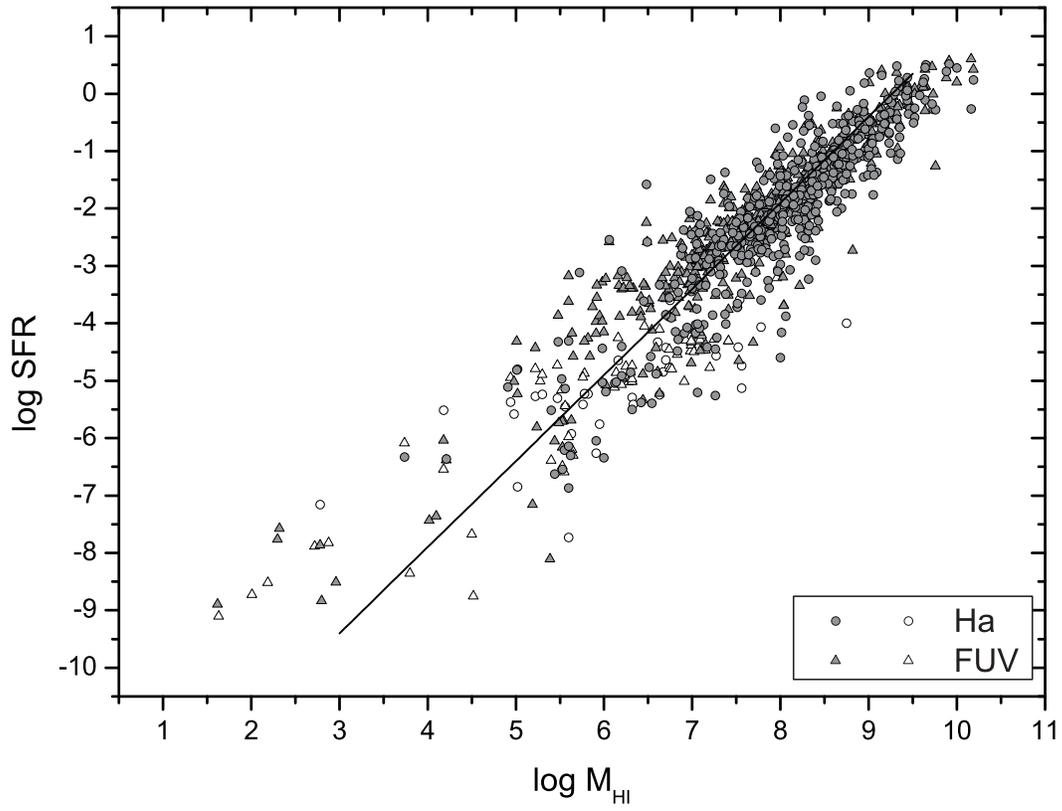}
\caption{Integral SFR vs. total hydrogen mass for the Local Volume galaxies.
           Indication of galaxies with different SFR sources are the same as
           in Fig.~\ref{Fig4}. The solid line traces the slope  3/2 corresponding to the
           Schmidt-Kennicutt law.}
\label{Fig5}
\end{figure*}

Fig.~\ref{Fig5} presents the distribution of Local Volume
galaxies according to their integral SFRs and total hydrogen
masses. As one can see, the galaxies of different morphological
types closely follow the known Schmidt-Kennicutt law with the
slope of 3/2.

\begin{figure*}
\centerline{
\begin{tabular}{c}\\[-20mm]
\includegraphics[width=0.9\textwidth,clip]{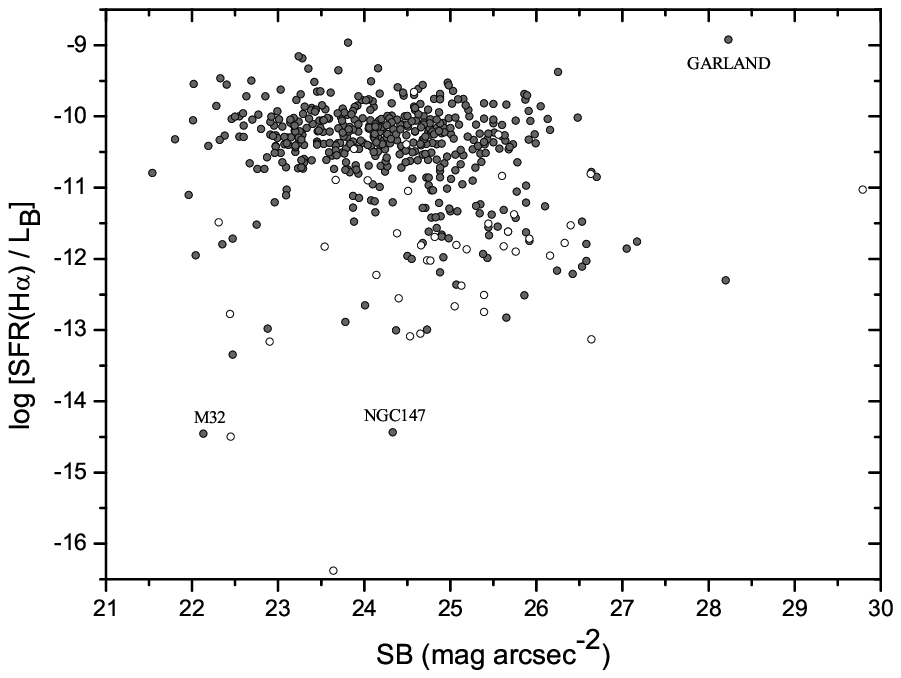}\\[-15mm]
\includegraphics[width=0.9\textwidth,clip]{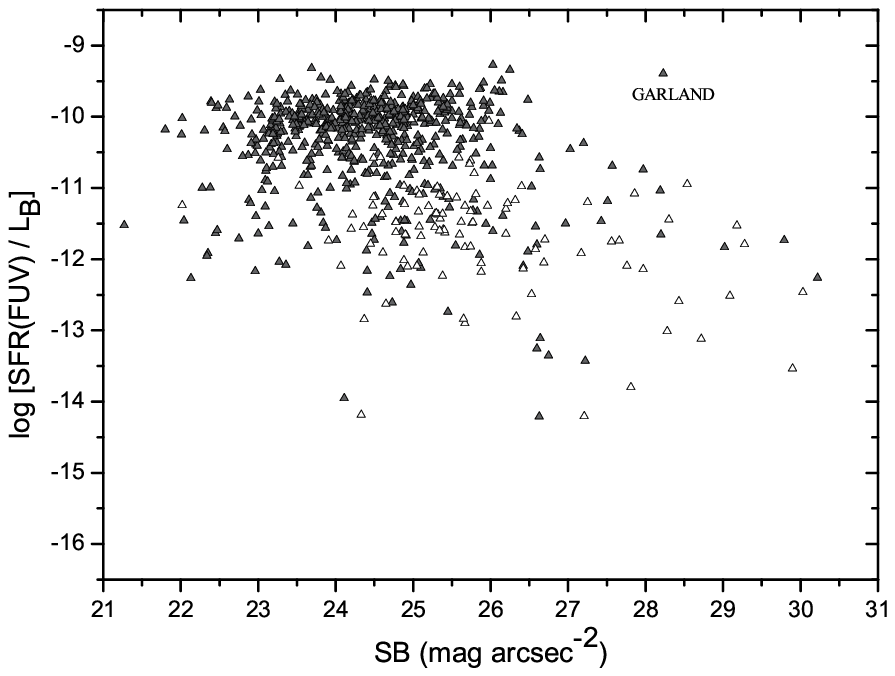}\\[-10mm]
\end{tabular}
}
\caption{Specific star formation rate of nearby galaxies vs. their blue
       surface brightness. The top panel is derived from the $H\alpha$ imaging data, and the bottom panel --- from
       FUV GALEX survey.}
\label{Fig6}
\end{figure*}

 One of the most important global parameters of galaxies is their mean optical surface brightness, SB.
 The galaxies in the Local Volume  have a scatter of mean   surface brightnesses in the B-band in the range of 21 to 30 mag arcsec$^{-2}$.
 Such enormous differences can significantly affect the efficiency and rate of star formation in the galaxies (Boissier et al. 2008,
 Meurer et al. 2009). The upper panel of Fig.~\ref{Fig6} shows the relative SFR    per
  $L_B$-luminosity unit of a given galaxy, determined by   its $H\alpha$ flux depending on the surface brightness. As follows from this diagram,
  the specific star formation rate is almost
not correlated with the surface brightness of the galaxy up to the
value of  SB $\simeq 26.5$ mag arcsec$^{-2}$. The  SFR estimates
by the FUV flux (the bottom panel of Fig.~6) extend into the
region of fainter surface brightnesses, where there is a tendency
of a declining specific star formation rate at SB$>26.5$ mag
arcsec$^{-2}$. A comparison of the upper and lower panels shows
that the upper limit of SFR/$L_B$ looks sharper for the FUV
fluxes.This feature can be explained by the fact that the
$H\alpha$ flux characterizes the activity of star formation on a
shorter time scale and hence its reaction   to the bursts of star
formation is less robust than that of the FUV flux.

 \section{SFR and dwarf morphology}

Around 75\% of the Local Volume population are dwarf galaxies with luminosities and sizes smaller than that of the Large Magellanic Cloud.
In the de Vaucouleurs morphological classification they correspond to two types: T = 9 (irregular magellanic = Im, blue compact dwarf = BCD)
and T = 10 (dwarf irregular = Ir).
 Moreover, dwarf elliptical galaxies --- dE and diffuse spheroidals --- Sph are usually added to the normal elliptical (E) and
 lenticular (S0) galaxies, being attributed the types  T$ <1$. The shortcomings of such a simplified classification of dwarf systems have
 become apparent, hence, van den Bergh (1959) proposed a more refined scheme where dwarf galaxies assumed to be vary by luminosity classes.

In the UNGC, we have used a two-dimensional classification of
dwarfs based on the available observational characteristics. In
the vertical
 direction of our scheme, presented in Table~\ref{t:class}, dwarf galaxies varied in surface brightness gradations, the   typical values of
 which amounted to:  23.0 --- High SB, 24.2 --- Normal SB, 25.2 --- Low SB and 27.6 --- eXtremely low SB, in units of [mag arcsec$^{-2}$].
 In the horizontal direction we have divided   dwarf
 galaxies into red (old), intermediate, and blue (young). The first category are spheroidal (Sph), dwarf lenticular (dS0) and
 dwarf elliptical (dE) galaxies, as well as intergalactic globular clusters (gc). The second group are the objects of mixed stellar population
 Sph/Ir (or Transition), as well as the dE and dS0 dwarfs with emission spectra.
The right column  combines the dwarf system of  BCD, Im, Ir types,
as well as the rare cases of intergalactic HI-clouds with no signs
of star formation. Such an approach allowed us to avoid
unpleasant cases when a dwarf galaxy of intermediate type could
jump over due to a misclassification from one end of de
Vaucouleurs sequence
 (T $ <1$) to another one (T = 9, 10).

The bold numbers in Table~\ref{t:class} show how the Local Volume
dwarfs are   distributed by the cells of our two-dimensional
scheme. The population of cells appears to be very uneven, which
is caused both by the natural causes and the effects of
observational selection.

  For the dwarf galaxies belonging to each cell of Table~\ref{t:class}, we have determined the average color indices
$\langle m_{FUV} - B\rangle$, $\langle B - m_{H\alpha}\rangle$ and
  $\langle B-m_{21}\rangle$, corrected for the Galactic and internal extinction. Here, the apparent magnitudes were expressed through the
  corresponding fluxes as

  $$m_{FUV}=23.90-2.5\log F_{FUV}[mJy],$$
  $$ m_{H\alpha} = -13.64 -2.5\log F_{H\alpha} [erg\cdot cm^{-2}\cdot s^{-1}],$$
  $$ m_{21} = 17.4-2.5\log F_{HI} [Jy\cdot km\cdot s^{-1}].$$

The average value for each color index, standard deviation and
the number of galaxies with the above attributes are  presented
in the cells of Table~\ref{t:aver}, which are identical to the
corresponding cells in Table~\ref{t:class}. An analysis of these
data reveals the following features.

In each subtype of surface brightnesses of dwarf galaxies where
there exists a sufficient statistics, the average color index
$\langle m_{FUV} -B\rangle$ increases towards blue  $\rightarrow$
mixed $\rightarrow$ red.   There is also a tendency of an
increasing color index dispersion in the same direction.

  In the transition from blue to red dwarfs, the average color indices  $\langle B -m_{H\alpha}\rangle$ and
   $\langle  B - m_{21}\rangle$ are systematically decreasing, which indicates a relative weakening of the emission  in the   $H\alpha$ and $HI$
   lines. The dispersion of the color indices tends to grow from  blue  to old red objects too.

The transition from dwarf galaxies of high surface brightness to
the low and extremely low surface brightness has little effect on
the average color index $\langle m_{FUV} -B\rangle$. For the red
dwarfs the value of  $\langle B -m_{H\alpha}\rangle$ increases,
while for the blue dwarfs it decreases towards the objects of low
surface brightness. The color index $\langle B - m_{21}\rangle$
grows from compact dwarfs to extremely diffuse dwarfs, indicating
that the   proportion of gas relative to the stellar mass is
greater in the latter galaxies.

   \section{Starbursts in low-mass galaxies}

The literature repeatedly expressed concerns that the conversion
of gas into stars in dwarf galaxies has a oscillatory, starburst
character   (Dohm-Palmer et al. 2002, Skillman 2005, McConnachie
et al. 2006, Karachentsev \& Kaisin 2007, McQuinn et al. 2009).
Stinson et al. (2007) performed a    numerical simulation of this
process and have shown that in gas-rich dwarf galaxies with
masses of  $\log(M/M_{\odot})<9$ the  bursts of star formation
can vary the SFRs   severalfold   on the typical time scale of
$\sim3 \cdot 10^8$ years. According to the models of periodic
SFH for dwarf galaxies, considered by Weisz et al. (2012), the
low-mass galaxies increase their SFRs by a factor of 10 -50 over
the average SFR during the inter-burst period. Their burst
duration and cycle periods are typically tens of Myr and 200 -
300 Myr, respectively. The same variability, determined by the
internal parameters of a dwarf galaxy itself  is superimposed by
some  external factors: the excitation of  star formation
activity due to the tidal effects from a massive neighboring
galaxy, as well as the  sweeping out gas from the dwarf system as
it passes through the dense regions of its massive neighbor. The
relative role of external and  internal factors, affecting the
evolution of a given dwarf galaxy should evidently depend on the
density of its environment.

\begin{figure*}
\centerline{
\begin{tabular}{c}\\[-25mm]
\includegraphics[width=0.9\textwidth,clip]{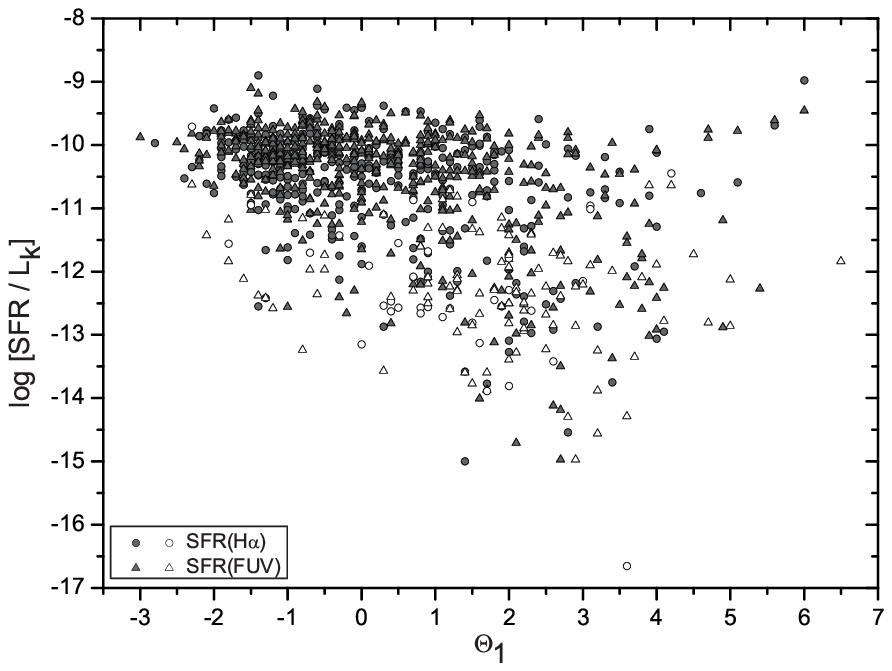}\\[-20mm]
\includegraphics[width=0.9\textwidth,clip]{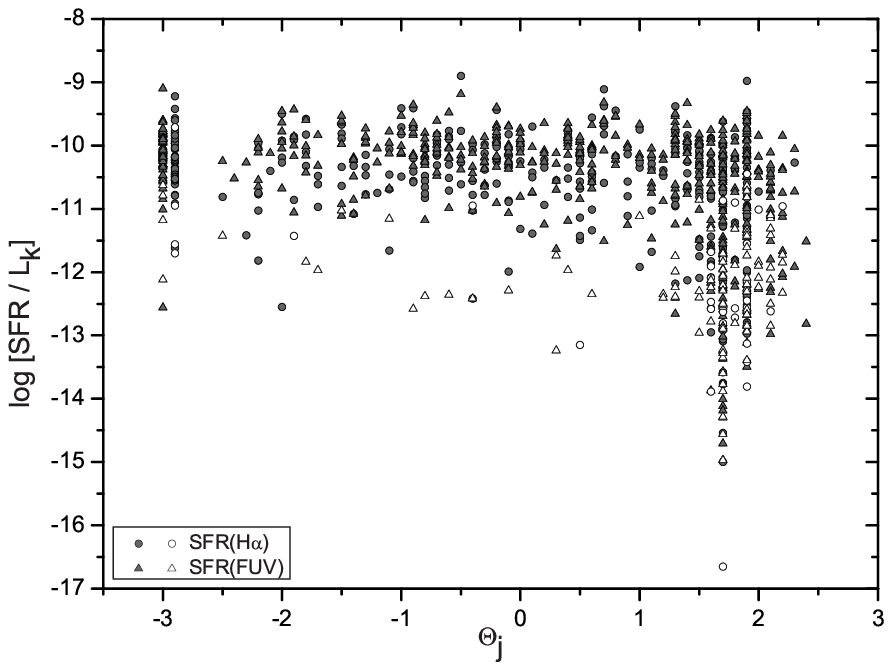}\\[-10mm]
\end{tabular}
} \caption{Specific star formation rate for the dwarf ($log M_* <
9$) Local Volume galaxies as a function of two environment
parameters: $\Theta_1$, determined by the nearest significant
neighbor (top panel), and $\Theta_j$, indicating
       the stellar mass density contrast within the 1 Mpc radius in the units of
       mean global density (bottom panel). Symbols are the same as in Fig.~\ref{Fig4}
       - Fig.~\ref{Fig6}. Negative values of $\Theta$ correspond to the field galaxies,
       while positive ones correspond to the group members.}
\label{Fig7}
\end{figure*}

Panels of  Fig.~\ref{Fig7} show the distribution of dwarf
galaxies with stellar masses  $\log(M_*/M_{\odot})<9$ by  the
SSFR   depending on their environment. The galaxies with SFR
estimated from the $H\alpha$ flux and FUV flux are depicted by the
circles and triangles, respectively. The upper limits of the flux
are shown by empty symbols. The horizontal axis of the top panel
represents the so-called "tidal index"  of a galaxy
$$\Theta_1=\max[\log(M_n/D_n^3)]-10.96, \,\,\,\, n=1, 2\ldots N  $$
  which is expressed in terms of  mass $M_n$ and spatial distance $D_n$ of the most significant neighboring galaxy (Karachentsev et al. 2013).
  Positive values of $\Theta_1$ correspond to the members of groups, and  negative --- to the field galaxies. The bottom panel uses another
  dimensionless indicator as an argument,
     $$\Theta_j=\log[j_K(1{\rm Mpc})/j_{K,global}],  $$
which characterizes the contrast of stellar density around the
considered galaxy in a sphere with a 1 Mpc radius, expressed in
the units
    of  global mean density
     $j_{K,global} = 4.28\cdot10^8L_{\odot}$ Mpc$^{-3}$ (Jones et al. 2006). For extremely isolated galaxies with no neighbors within 1 Mpc,
     the values of  $\Theta_j$ are formally accepted by us to amount  to --3.

As follows from the data presented, the smallest scatter in the specific star formation rate is found in isolated dwarf galaxies. In
high density regions, $\Theta_1>0$ or $\Theta_j>1$,  there is a noticeable amount of dwarf objects with depressed star formation rates
$\log(SFR/M_*) < -11.5$. Their relative number
   is not too significant, only  $\sim 15$\%,  but this figure may be seriously biased by the effects of observational selection.
   Note also that at the level of highest SSFR values, we practically do not recognize any dwarf galaxies, in which an increase
   in the SFR would be triggered by a dense  neighborhood.
  A rare and outstanding example of this is \ "Garland" \ --- a dwarf galaxy at the NGC~3077 periphery, which looks like a chain of emission
  HII-regions, immersed in a cloud of molecular hydrogen (Karachentsev et al. 1985, Walter et al. 2002).

Globally, the data presented in  Fig.~\ref{Fig7} show that in the
majority of dwarf galaxies, the rates of transformation of gas
into stars are largely determined by the internal processes rather
than any external effects.

   To characterize the evolutionary status of a sample of galaxies, Karachentsev \& Kaisin (2007) proposed to use a diagnostic "past-future" (=PF)
   diagram, where the dimensionless parameters
   $$P = \log(SFR\cdot T_0/L_K), $$
     $$F = \log(1.85\cdot M_{HI}/SFR\cdot T_0) $$
are independent of errors in finding distances to the galaxies.
The parameter $P$ is actually the specific star formation rate
over the entire
  age scale of the universe, $T_0=13.7$ Gyr.  The  $F$ parameter corresponds to the notion of  gas depletion time, expressed in the units of $T_0$.
  The coefficient of 1.85 at $M_{HI}$ is introduced in order to account for the contribution of helium and molecular hydrogen in the total mass
  of gas (Fukugita \& Peebles 2004).

 The summary of observational data on the integral SFR (columns 5 and 8) and evolutionary parameters
 $\{P, F\}$  (columns 6,7 and 9,10) for 802 galaxies of the  Local Volume is given in Table~\ref{t:sfr}. The first two columns 
 contain the name of the galaxy and its equatorial coordinates, the third column indicates the galaxy morphology by de Vaucouleurs scale,
 and the fourth column lists the B-band absolute magnitude corrected for the Galactic and internal extinction.

\begin{figure*}
\centerline{
\begin{tabular}{c}\\[-25mm]
\includegraphics[width=0.5\textwidth,clip]{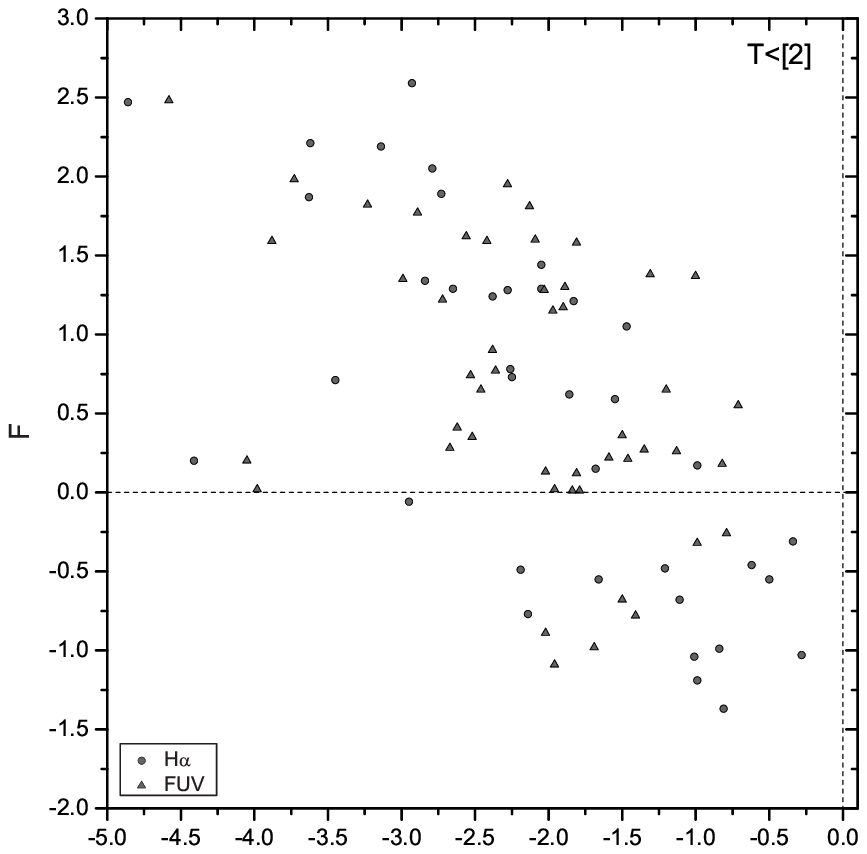}\\[-15mm]
\includegraphics[width=0.5\textwidth,clip]{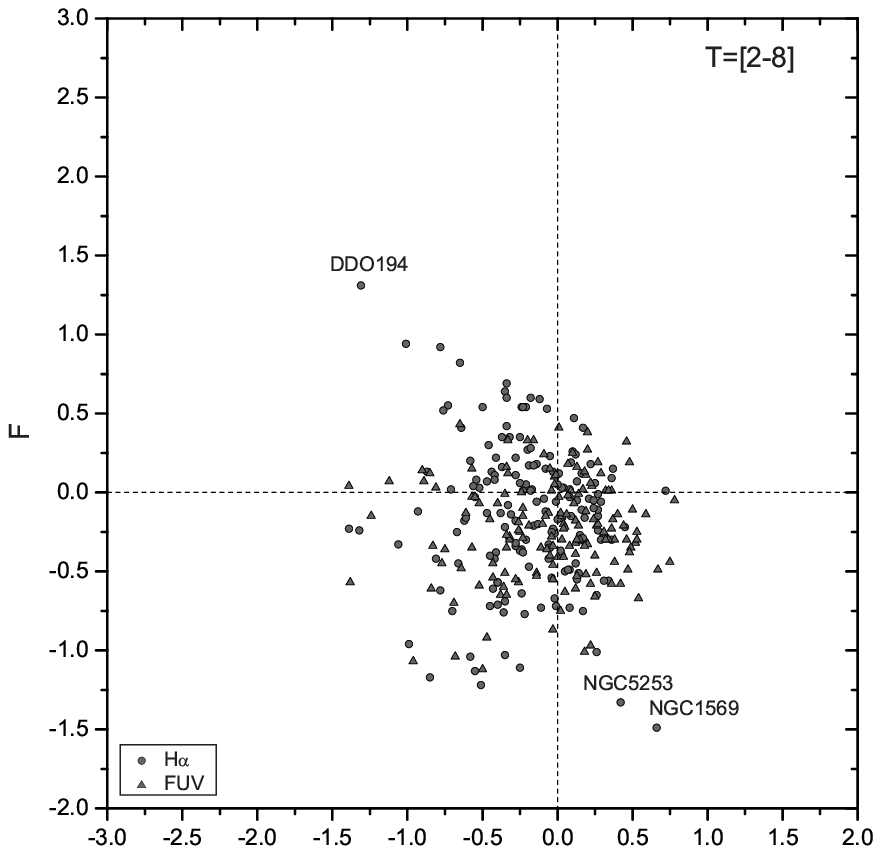}\\[-15mm]
\includegraphics[width=0.5\textwidth,clip]{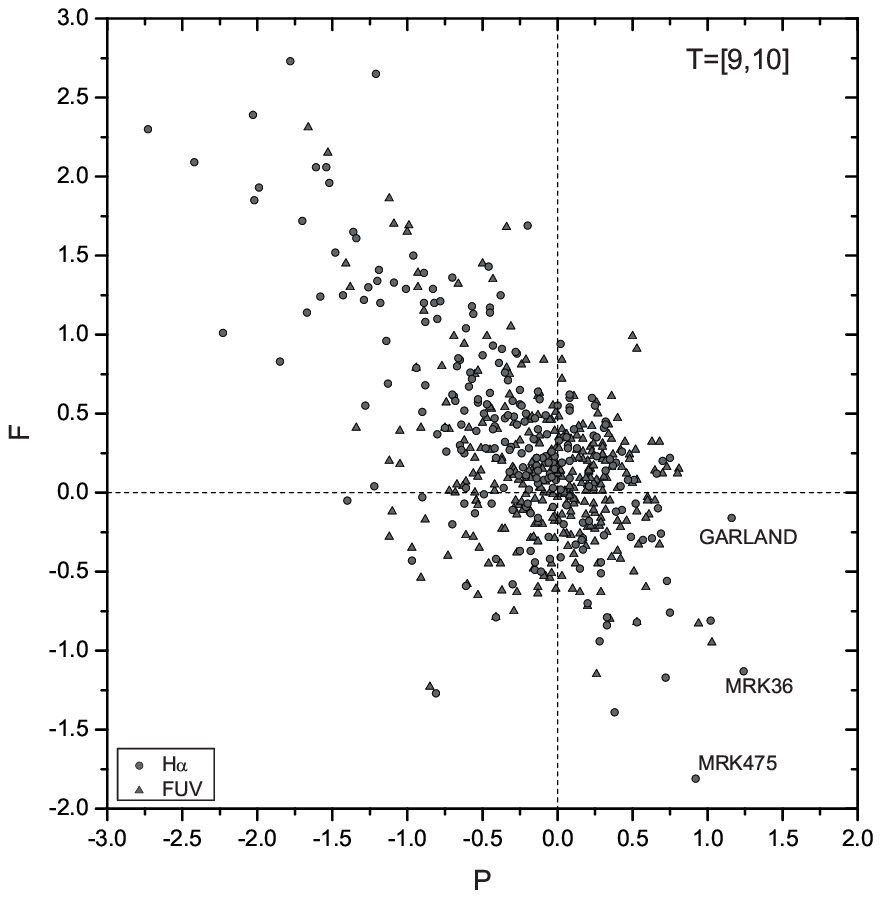}\\[-10mm]
\end{tabular}
} \caption{The diagnostic  "past-future" diagram for the
early-type (E,S0, dSph)
       galaxies (top panel), spiral galaxies (middle panel), and late-type
       (BCD, Im, Ir) dwarfs (bottom panel).}
\label{Fig8}
\end{figure*}

Figure~\ref{Fig8} reproduces the diagnostic diagrams of   $\{P,
F\}$ for the early-type  T$<2$  galaxies of the Local Volume (the
top panel), for the spiral  galaxy types T = (2--8) (middle panel)
and dwarf galaxy types T=(9,10) (the bottom panel).

The top panel shows that in the E and S0 galaxies, the current
star formation rates are by 2--3 orders of magnitude lower than
the earlier rates, which formed the observed stellar mass of
these galaxies. Given the present  reserves of gas, their
observed  \ "glowing" \ star formation rates can be maintained at
the average  on the scale of several more Hubble times.

As  follows from the middle panel, a typical spiral galaxy has
time to reproduce  its stellar mass at its observing  SFR. The
gas  reserves in a typical spiral are enough to keep the current
rate of gas conversion into stars on a scale of   $\sim 10$ Gyr.
In other words, the disks of galaxies act as rhythmic stellar
factories and   are yet   located half way of their evolution.
The farthest distance from the origin    ${P=0, F=0}$  is shown
by the   galaxy NGC~1569, which reveals a burst of  star
formation in the central region and   radial expansion of
emission filaments (Israel 1988, Hodge 1974).

 The data on the bottom panel demonstrate   that the majority of dwarf Ir, Im, BCD galaxies, the same as the spirals, reproduce their
 stellar mass at the  current SFR values. The gas reserves in a typical representative of this population are sufficient to maintain
 the average observed SFR level for several more Hubble times. In this sense, the secular evolution of the late-type dwarfs can be
 characterized as inhibited, or "lethargic" evolution.

   Apart from the main concentration of dwarf galaxies near the origin ${P=0, F=0}$, about a quarter of the population of
   {Ir, Im, BCD}-dwarfs reveal  an elongation along the diagonal of  $F=-P$.  This effect is easily explained by the assumption
   that a part   of the dwarf population is in a starburst
   state, which is then followed by a more prolonged star formation depression stage. Statistics of starbursting dwarfs in the
   right bottom corner of the bottom panel tells us that only $\sim1/20$ part of the total amount of dwarf galaxies are experiencing
   vigorous bursts. This result is fully consistent with the conclusion by Lee et al. (2009) that \ "dwarfs that are currently
   experiencing massive global bursts are just the $\sim6$\% tip of a low-mass galaxy iceberg".
   Table~\ref{t:active} lists  15 most representative
   starburst- state galaxies.  Its columns contain: (1) --- the name of the active galaxy, (2-3) --- the P parameter obtained from the $H\alpha$ and FUV fluxes,
  (4) --- morphology, (5) ---   absolute B-magnitude, (6) --- tidal index.
  The most extreme P values
  are observed in dwarfs with SFR estimates from the $H\alpha$ flux. This situation fits into the general concept of the recurrent starburst
  activity of low-mass galaxies, since the  $H\alpha$ fluxes fix the SFR value at a shorter term $(\sim10^7$ Myr), than the
  FUV fluxes do    ($\sim10^8$ Myr).

The data of Table~\ref{t:active} again show that the starbursts
depend little on the galaxy neighborhood density.

\section{Concluding remarks}

   We have discussed the available observational data on the current rate of star formation in the galaxies with distances of
   $D<11$ Mpc, which was determined from the integral flux of galaxy in the emission  $H\alpha$ line or from the FUV flux, obtained
   at the GALEX orbital telescope. Our sample of galaxies has two advantages: 1) it is the most representative of all the existing
   samples in the Local Volume, 2) at its formation we have not used  any restrictions on the morphological properties  of galaxies.
   About 3/4  of the sample are dwarf galaxies that we have classified by the gradations of surface brightness and color.

The population of dwarf galaxies shows signs of stochastic
starburst activity, which scatters them on the diagnostic $\{P,
F\}$ diagram  along the diagonal of  $F=-P$.  On the average,
blue dwarf galaxies (Ir, Im, BCD) possess current  star formation
rates, sufficient   to regenerate their observed stellar masses,
and their gas reserves allow them to maintain the observed
average SFR  on the scale  of several Hubble times. This statement
does not apply to the dwarf spheroidal companions that have lost
their gas component passing
   through the dense regions of massive neighboring galaxies.

Spiral galaxies of  Sa--Sm  types (T=2--8)   on average have
about the same specific  rate of star formation as  the  Ir, Im
and BCD dwarf systems. Their dispersion in the  $\{P, F\}$
diagram   is much smaller than that of dwarfs. Apparently, the
disks of spiral
 galaxies convert  gas into stars in a regular fashion, which is determined by  purely internal mechanisms with a small influence of
 the external environment.

For the Local Volume galaxies of different morphological
types there exists  an upper limit of the specific star formation
rate, $\log (SFR/M_*)_{lim}\simeq-9.4$   [yr$^{-1}]$,  to which
corresponds the dimensionless parameter  $P_{lim}\simeq 0.75.$
 Above this limit, there are only a few  $(\sim1$\%)
low-mass galaxies that fall into this region in the state of a
short vigorous starburst, or due to uncertainty in the
estimates of their SFR and P (compare the data on P in columns
(2) and (3) of Table~\ref{t:active}). The presence of this $\log (SFR/M_*)_{lim} = -9.4$ 
limit is well seen in Fig.9 by Schiminovich et al. (2010) for a sample of 190 massive 
HI- selected galaxies from Arecibo SDSS \& GALEX survey. Recently, just the same
upper limit was also found for a sample of 500 local isolated
galaxies and a sample of 270 nearby Markarian galaxies
(Karachentsev et al. 2013b). The derived upper limit in specific
star formation rate deserves attention of theoreticians. It may
indicate that the conversion of gas into stars is regulated by a
rather rigid feedback, when an excessively  active star
formation  rate is blocked by the depletion of internal resources
for it.

 The population of E, S0 and dSph galaxies in the Local Volume is characterized by very low rates of star formation. To reproduce the observed
   stellar mass of these galaxies, their average SFR in the past should have been   tens to hundreds of times higher than  the one observed
   now. It should be noted, however, that these arguments suggest the evolution of galaxies within the  \ "closed box" scheme. There are ample
   evidence that galaxies  increase their mass with time by the accretion of baryonic matter from the intergalactic space (Marinacci et al. 2010,
Cattaneo et al. 2011). The presence of the isolated E and S0
galaxies:   NGC~404, NGC~855 and NGC~4600 with active emission
nuclei (Moiseev et al. 2010) in the Local Volume may serve as a
confirmation of the fact that the process of accretion of
intergalactic baryons was at work not solely a  long time ago,
but it also extends into the present epoch.

\acknowledgments

      The authors are grateful to the referee for the comments that have improved the paper.
      We thank Dmitry Makarov  for the useful discussions. This work was supported in part by the RFBR grants no.~12--02--91338 and 13--02--92690.
      We acknowledge the support for proposals GO 12546, 12877, 12878, which was provided by the NASA through grants of the Space Telescope
      Science Institute.
      This research has made use of the survey web services of the NASA/IRAC Extragalactic Database (NED) and the Galaxy Evolution Explorer (GALEX).

{}
\clearpage

\begin{table}
\caption{Classification for dwarf galaxies}
\begin{tabular}{ll|c|c|c|} \hline
&{\bf H}igh & {\bf 2} & {\bf 1} & {\bf 13} \\
&& dE, gc & dEem & BCD \\
\hline
&{\bf N}ormal & {\bf 23} & {\bf 15} & {\bf 320} \\
&& dS0,Sph & dS0em,Tr & BCD,Im,Ir \\
\hline
&{\bf L}ow &{\bf 60} & {\bf 43} & {\bf 115}\\
&& Sph& Ir/Sph,Tr & Ir\\
\hline
&{\bf X}-Low & {\bf 49} & {\bf 4}& {\bf 6}\\
&& Sph & Tr & Ir,HI cld \\
\hline
&{\bf $\uparrow$}& {\bf Red} & {\bf Mixed} & {\bf Blue} \\
&{\bf SB}& {\bf Gas content $\longrightarrow$} & & {\bf $\longleftarrow$ Color Index}\\
 \end{tabular}
 \label{t:class}
\end{table}

\begin{table}
  \caption{Average colors for different  dwarf types}
  \begin{tabular}{|c|c|rrr|rrr|rrr|} \hline
  & & & & & & & & & & \\
  Color & SB &
  \multicolumn{3}{c|}{Red} &
  \multicolumn{3}{c|}{Mixed} &
  \multicolumn{3}{c|}{Blue} \\

  & &$\langle \,\,\,\,\,\, \rangle$ & $\sigma$& n & $\langle \,\,\,\,\,\, \rangle$ & $\sigma$& n& $\langle\,\,\,\,\,\,  \rangle$ & $\sigma$& n\\
  \hline
              & H & 4.93 & 0.35 & 2  & $-$  & $-$  & 1  & 1.57 & 0.79 & 9 \\
  $m_{FUV} -B$& N & 6.07 & 1.90 & 17 & 4.41 & 1.34 & 14 & 1.72 & 0.96 & 260\\
              & L & 5.64 & 1.63 & 51 & 4.84 & 1.52 & 42 & 2.26 & 1.41 & 87 \\
              & X & 5.98 & 2.73 & 44 & 4.53 & 3.02 & 3  & 1.55 & 1.58 & 2 \\
              \hline
                   & H & $-$  & $-$  & 0  & $-$  & $-$  & 1  & $-$2.96 & 1.45 & 5 \\
 $B - m_{H\alpha}$ & N & $-$11.07 & 5.15 & 5& $-$5.84 & 2.43 & 7 & $-$4.31 & 1.48 & 153\\
                   & L & $-$8.36 & 1.69 & 21 & $-$7.80 & 1.08 & 12 & $-$5.19 & 1.90 & 50 \\
                   & X & $-$7.75& 1.93& 6 & $-$  & $-$  & 0 & $-$  & $-$  & 1 \\
  \hline
                   & H & $-$  & $-$  & 1 & $-$  & $-$  & 1 & $-$2.17 & 0.97 & 9 \\
 $B - m_{21}$      & N & $-$3.23 & 3.84 & 8 & $-$2.19 & 1.16 & 12 & $-$0.79 & 1.11 & 230 \\
                   & L & $-$2.20 & 1.94 & 31 & $-$1.68 & 1.51 & 30 & 0.01 & 1.05 & 78 \\
                   & X & $-$2.34 & 3.18 & 26 & $-$  & $-$  & 1 & 0.17 & 1.46 & 2 \\
                   \hline
 \end{tabular}
 \label{t:aver}
 \end {table}

 \hoffset=-2.0cm   
\tiny{
\begin{deluxetable}{lcrrrrrrrr}
\tablecolumns{10}
\tablewidth{0pc}
\tablecaption{List of the nearby galaxies with measured SFR}
\tablehead{
\multicolumn{1}{c}{Galaxy}&
\multicolumn{1}{c}{RA (J2000.0) Dec.} & 
\multicolumn{1}{c}{T} & 
\multicolumn{1}{c}{$M_B$}&
\multicolumn{1}{c}{$SFR_{H\alpha}$} & 
\multicolumn{1}{c}{$P_{H\alpha}$}& 
\multicolumn{1}{c}{$F_{H\alpha}$} & 
\multicolumn{1}{c}{$ SFR_{FUV}$} & 
\multicolumn{1}{c}{$P_{FUV}$} & 
\multicolumn{1}{c}{$F_{FUV}$}  \\
\colhead{1}&
\colhead{2}&
\colhead{3}&
\colhead{4}&
\colhead{5}&
\colhead{6}&
\colhead{7}&
\colhead{8}&
\colhead{9}&
\colhead{10}}
\startdata
 UGC12894          & 000022.5$+$392944 & 10 &  $-$13.31 &  $-$2.48     &  0.08 &  0.53 &  $-$2.03      &   0.53 &   0.08 \\
 WLM               & 000158.1$-$152740 &  9 &  $-$14.06 &  $-$2.69     & $-$0.24 &  0.65 &  $-$2.24      &   0.21 &   0.20 \\
 And XVIII         & 000214.5$+$450520 &$-$3&  $-$9.11  &           &       &       & $<-$5.81      &  $<-$2.27 &        \\
 ESO409-015        & 000531.8$-$280553 &  9 &  $-$14.35 &  $-$1.57     &  0.57 & $-$0.31 &  $-$1.82      &   0.32 &  $-$0.06 \\
 AGC748778         & 000634.4$+$153039 & 10 &  $-$10.04 &            &       &       &  $-$3.65      &   0.22 &   0.29 \\
 And XX            & 000730.7$+$350756 &$-$3&  $-$5.77  &           &       &       &  $-$5.96      &  $-$1.08 &        \\
 UGC00064          & 000744.0$+$405232 & 10 &  $-$14.75 &            &       &       &  $-$1.63      &   0.35 &   0.34 \\
 ESO349-031        & 000813.3$-$343442 & 10 &  $-$11.87 &  $-$4.03     & $-$1.01 &  1.29 &  $-$3.02      &   0.00 &   0.28 \\
 NGC0024           & 000956.4$-$245748 &  5 &  $-$18.32 &  $-$0.74     & $-$0.35 & $-$0.22 &  $-$0.34      &   0.05 &  $-$0.62 \\
 NGC0045           & 001403.9$-$231056 &  8 &  $-$18.53 &  $-$0.26     &  0.27 & $-$0.14 &  $-$0.17      &   0.36 &  $-$0.23 \\
 NGC0055           & 001508.5$-$391313 &  8 &  $-$18.41 &  $-$0.35     &  0.30 & $-$0.07 &  $-$0.21      &   0.44 &  $-$0.21 \\
 NGC0059           & 001525.1$-$212638 &$-$3&  $-$15.74 &  $-$1.91     & $-$0.50 & $-$0.56 &  $-$2.20      &  $-$0.79 &  $-$0.27 \\
 ESO410-005        & 001531.4$-$321048 & 10 &  $-$11.58 & $<-$6.26     & $<-$3.01 & $>$2.30 &  $-$3.97      &  $-$0.72 &   0.01 \\
 And XIX           & 001932.1$+$350237 &$-$3&  $-$8.32  &           &       &       & $<-$6.27      &  $<-$2.41 &        \\
 IC0010            & 002024.5$+$591730 & 10 &  $-$15.99 & $-$1.54     &  0.12 & $-$0.33 &             &        &        \\
 And XXVI          & 002345.6$+$475458 &$-$3&  $-$6.47  &            &       &       & $<-$6.30      &  $<-$1.70 &        \\
 Sc22              & 002351.7$-$244218 &$-$3&  $-$10.46 & $<-$5.58     & $<-$2.58 &  $>$0.69 &  $-$5.02      &  $-$2.02 &   0.13 \\
 Cetus             & 002611.0$-$110240 &$-1$&  $-$10.18 & $<-$5.51     & $<-$2.40 &$>-$0.18 & $<-$6.54      &  $<-$3.43 &   $>$0.85 \\
 ESO294-010        & 002633.3$-$415120 & 10 &  $-$10.91 & $-$4.32     & $-$0.43 &$-$0.07 & $-$3.86      &   0.03 &  $-$0.53 \\
 UGC00288          & 002904.0$+$432554 & 10 &  $-$13.83 & $-$2.62     & $-$0.27 &  0.43 &  $-$2.34      &   0.01 &   0.15 \\
 \hline							   		 		    
 \multicolumn{10}{l}{\small{\textbf{Note.} Only a portion of this table is shown here to demonstrate its form and content.}}\\
\multicolumn{10}{l}{\small{     Machine-readable version of the full table is available.}}\\
\enddata
 \label{t:sfr}
\end{deluxetable}
}

 \begin{table}
  \caption{The most active galaxies in the Local Volume} 
  \begin{tabular}{lcclcr} \hline
   
   Name & $P_{H\alpha}$ & $P_{FUV}$&  T & $M_B$ & $\Theta_1$\\
   (1) & (2) & (3) & (4) & (5) & (6) \\
   \hline
   Mrk~36 & 1.24& 0.94& BCD   & $-$14.04 &$-$1.4 \\
   NGC1592& $-$    & 1.03& Ir  & $-$15.50 &$-$1.5 \\
   Garland& 1.15& 0.68& Ir    & $-$11.40 & 6.0 \\
   DDO~169NW & $-$ & 0.80 & Ir   & $-$10.16 & 0.0 \\
   Mrk~209& 1.02& 0.54& Ir    & $-$13.69 & $-$0.6 \\
   UGCA~292& 0.75 & 0.81& Ir    & $-$11.79 & $-$0.6 \\
   NGC4597 & 0.73& 0.78& Sdm   & $-$17.81 & 0.0 \\
   NGC5408 & 0.73 & $-$ &Im      & $-$16.51 &$-$0.2 \\
   DDO~143& 0.70 & 0.73 & Ir  & $-$13.88 & $-$0.5 \\
   UGC4483 & 0.67& 0.65 & Ir   & $-$12.73 & 0.6 \\
   GR~8& 0.63 &0.68 & Ir      & $-$11.96 & $-$1.4 \\
   UGC6456& 0.70 & 0.60& Ir    & $-$14.08 & $-$0.1 \\
   DDO~53 & 0.66& 0.60& Ir    & $-$13.37 & 0.8 \\
   NGC4861&0.76 & 0.43 & Im        &$-$16.52  & 0.3 \\
   Mrk~475& 0.92 & 0.26& BCD  & $-$13.46 & $-$1.2 \\
   \hline
   \end{tabular}
   \label{t:active}
   \end{table}


\begin{thebibliography}{}

\bibitem{Bell2003} Bell E.F., McIntosh D.H., Katz N.,\& Weinberg M.D., 2003, ApJS, 149, 289
\bibitem{Boissier2008} Boissier S.,Gil de Paz A., Boselli A., et al. 2008, ApJ, 681, 244
\bibitem{Bouchard2009} Bouchard A., Da Costa G.S.,\& Jerjen H.,  2009  AJ, 137, 3038
\bibitem{Cattaneo2011} Cattaneo A., Mamon G.A., Warnick K.,\& Knebe A., 2011, A \& A, 533, 5
\bibitem{Dohm2002} Dohm-Palmer R.C., Skillman E.D., Mateo M., et al. 2002, AJ, 123, 813
\bibitem{Epinat2008} Epinat B., Amram P.,\& Marcelin M., 2008, MNRAS, 390, 466
\bibitem{Fukugita2004} Fukugita M.,\& Peebles P.J.E., 2004, ApJ, 616, 643
\bibitem{Fumagalli2011} Fumagalli M., Da Silva R.L. \& Krumholz M.R., 2011, ApJL, 741, L25 
\bibitem{Gil2003} Gil de Paz A., Madore B.F. \& Pevunova O. 2003, ApJS, 147, 29
\bibitem{Gil2007} Gil de Paz A., Boissier S., Madore B.F. et al. 2007, ApJS, 173, 185
\bibitem{Hodge1974} Hodge P.W., 1974, ApJ, 191L, 21
\bibitem{Huang2012} Huang S., Haynes M.P., Giovanelli R.,\& Brinchmann J., 2012, ApJ, 756, 113
\bibitem{Hunter2010} Hunter D.A., Elmegreen B.G.,\& Ludka B.C., 2010, AJ, 139, 447
\bibitem{Hunter2004} Hunter D.A. \& Elmegreen B.G. 2004, AJ, 128, 2170
\bibitem{Israel1988} Israel F.P., 1988, A\& A, 194, 241
\bibitem{James2004} James, P. A., Shane, N. S., Beckman, J. E. et al., 2004, A \& A, 414, 23
\bibitem{James2008} James P.A., Knapen J.H., Shane N.S. et al., 2008, A \& A, 482, 507
\bibitem{Jones2006} Jones D.H., Peterson B.A., Colless M.,\& Saunders,W., 2006, MNRAS, 369, 25
\bibitem{Kaisin2007} Kaisin  S.S., Kasparova A.V., Kniazev A.Yu.,\& Karachentsev I.D., 2007, Astron. Lett., 33, 1
\bibitem{Kaisin2006} Kaisin  S.S.,\& Karachentsev  I.D., 2006, Astrofizika, 49, 337
\bibitem{Kaisin2008} Kaisin  S.S.,\& Karachentsev  I.D., 2008, A\&A, 479, 603
\bibitem{Kaisin2011} Kaisin  S.S., Karachentsev  I.D.,\& Kaisina E.I., 2011, Astrofizika, 54, 353
\bibitem{Kaisina2012} Kaisina E.I., Makarov D.I., Karachentsev I.D.,\& Kaisin  S.S., 2012, AstBul, 67, 115
\bibitem{Karachentsev1985} Karachentsev I.D., Karachentseva V.E.,\& Borngen F., 1985, MNRAS, 217, 731
\bibitem{Karachentsev2005} Karachentsev I.D., Kaisin S.S., Tsvetanov Z.,\& Ford H., 2005, A\&A, 434, 935
\bibitem{Karachentsev2007} Karachentsev I.D.,\& Kaisin S.S., 2007, AJ, 133, 1883
\bibitem{Karachentsev2010} Karachentsev I.D.,\& Kaisin S.S., 2010, AJ, 140, 1241
\bibitem{Karachentsev2011} Karachentsev I.D., Kaisina E.I., Kaisin S.S.,\& Makarova L.N., 2011, MNRAS, 415L, 31
\bibitem{Karachentsev2013} Karachentsev I.D., Makarov D.I.,\& Kaisina E.I., 2013a, AJ, 145, 101
\bibitem{Karachentsev2013b} Karachentsev I.D., Karachentseva V.E., \& Melnyk O.V., 2013b, Astroph Bull, 68 (accepted)
\bibitem{Kennicut2012} Kennicutt R.C. \& Evans N.J., 2012, ARAA, 50, 531 
\bibitem{Kennicut1998} Kennicutt R.C., 1998, ARA\&A, 36, 189
\bibitem{Kennicut2008} Kennicutt R.C.,  Lee J.C.,  Funes J.G. et al., 2008, ApJS, 178, 247
\bibitem{Lee2011} Lee J.C., Gil de Paz A., Kennicutt R.C., et al. 2011, ApJS, 192, 6
\bibitem{Lee2009} Lee J.C., Kennicutt R.C.,  Funes J.G. et al.,  2009, ApJ,  692, 1305
\bibitem{Marinacci2010} Marinacci F., Binney J., Flaternali F., et al. 2010, MNRAS, 404, 1464
\bibitem{McConnachi2006} McConnachie A.W., Arimoto N., Irwin M.,\& Tolstoy E., 2006, MNRAS, 373, 715
\bibitem{McQuinn2006} McQuinn K.B., Skillman E.D., Cannon J.M., et al. 2009, ApJ, 695, 561
\bibitem{Meurer2009} Meuer G.R., Wong O.I., Kim J.H. et al. 2009, ApJ, 695, 765 
\bibitem{Meurer2006} Meurer G.R., Hanish D.J., Ferguson H.C., et al. 2006, ApJS, 165, 307
\bibitem{Moiseev2010} Moiseev A.V., Karachentsev I.D.,\& Kaisin S.S., 2010, MNRAS, 403, 1849
\bibitem{Pflamm2009} Pflamm-Altenburg J., Weidner C.,\& Kroupa P., 2009, MNRAS, 395, 394
\bibitem{Pflamm2007} Pflamm-Altenburg J., Weidner C.,\& Kroupa P., 2007,ApJ, 671, 1550
\bibitem{Relano2012} Relano M., Kennicutt R.C., Eldridge J.J., et al., 2012, MNRAS, 423, 2933
\bibitem{Schiminovich2010} Schiminovich D., Catinella B., Kauffmann G. et al. 2010, 408, 919 
\bibitem{Schlegel1998} Schlegel D.J., Finkbeiner D.P., \& Davis M., 1998, ApJ, 500, 525
\bibitem{Skillman2005} Skillman E.D., 2005, New Astronomy survey, 49, 453
\bibitem{Stinson2007} Stinson G.S., Dalcanton J.J., Quinn T., et al. 2007, ApJ, 667, 170
\bibitem{Tully2008} Tully R.B., Shaya E.J., Karachentsev I.D. et al., 2008, ApJ, 676, 184
\bibitem{Tully1977} Tully R.B.,\& Fisher R.J., 1977, A \& A, 54, 661
\bibitem{Bergh1959} van den Bergh S., 1959, Publ. David Dunlap Obs., 2, 14
\bibitem{Verheijen2001} Verheijen, M.A.W., 2001, ApJ, 563, 694
\bibitem{Walter2002} Walter F., Weiss A., Martin C.,\& Scoville N., 2002, AJ, 123, 225
\bibitem{Weisz2012} Weisz D.R., Johnson B.D., Johnson L.C., et al. 2012, ApJ, 744, 44
\bibitem{Young1996} Young J.S., Allen L., Kenny J.D. et al., 1996, AJ, 112, 1903
\end{thebibliography}
  \end{document}